\documentclass[iop]{emulateapj}
\usepackage{hyperref}
\usepackage{ulem}
\usepackage{verbatim}
\usepackage{color}

\shorttitle{XMM-Newton observation of PSR B1133+16}
\shortauthors{Szary et al.}

\begin{document}

    \title{XMM-Newton observation of the nearby pulsar B1133+16.}

    \author{Andrzej Szary\altaffilmark{1,2}, Janusz Gil\altaffilmark{1}, Bing Zhang\altaffilmark{3}, Frank Haberl\altaffilmark{4}, George I. Melikidze\altaffilmark{1,5}, Ulrich Geppert\altaffilmark{1,6}, Dipanjan Mitra\altaffilmark{7, 1, 8}, Ren-Xin Xu\altaffilmark{9}}

    \email{aszary@astro.ia.uz.zgora.pl}

    \altaffiltext{1}{Janusz Gil Institute of Astronomy, University of Zielona G\'ora, Szafrana 2, 65-516 Zielona G\'ora, Poland}
    \altaffiltext{2}{ASTRON, the Netherlands Institute for Radio Astronomy, Postbus 2, 7990 AA, Dwingeloo, The Netherlands}
    \altaffiltext{3}{Department of Physics and Astronomy, University of Nevada Las Vegas, NV 89154, USA, zhang@physics.unlv.edu}
    \altaffiltext{4}{Max-Planck-Institut  f{\"u}r extraterrestrische Physik, Giessenbachstra{\ss}e, 85748 Garching, Germany}
    \altaffiltext{5}{Abastumani Astrophysical Observatory, Ilia State University, 3-5 Cholokashvili Ave., Tbilisi, 0160, Georgia}
    \altaffiltext{6}{German Aerospace Center, Institute for Space Systems, Robert-Hooke-Str. 7, 28359 Bremen, Germany}
    \altaffiltext{7}{National Centre for Radio Astrophysics, Ganeshkhind, Pune 411 007, India}
    \altaffiltext{8}{Physics Department, University of Vermont, Burlington, VT 05405}
    \altaffiltext{9}{School of Physics and Kavli Institute for Astronomy and Astrophysics, Peking University, Beijing 100871, China}

  \begin{abstract}
We constrain the X-ray properties of the nearby $(360\,{\rm pc})$, old ($5\,{\rm Myr}$) pulsar B1133+16 with $\sim   100\,{\rm ks}$ effective exposure time by {\it XMM-Newton}. 
The observed pulsar flux in the $0.2-3\,{\rm keV}$ energy range is $\sim 10^{-14} \, {\rm erg \, cm}^{-2} \, {\rm s}^{-1}$, which results in the recording of $\sim 600$ source counts with the EPIC pn and MOS detectors. 
The X-ray radiation is dominated by nonthermal radiation and is well described by both a single power-law model (PL) and a sum of blackbody and power-law emission (BB+PL).
The BB+PL model results in a spectral photon index $\Gamma=2.4^{+0.4}_{-0.3}$ and a nonthermal flux in the $0.2-3\,{\rm keV}$ energy range of $(7\pm 2) \times 10^{-15}\, {\rm erg \, cm}^{-2} \, {\rm s}^{-1}$. 
The thermal emission is consistent with the blackbody emission from a small hot spot with a radius of $R_{\rm pc} \approx 14^{+7}_{-5} \, {\rm m}$ and a temperature of $T_{\rm s} = 2.9^{+0.6}_{-0.4} \, {\rm MK}$. 
Assuming that the hot spot corresponds to the polar cap of the pulsar, we can use the magnetic flux conservation law to estimate the magnetic field at the surface $B_{\rm s} \approx 3.9 \times 10^{14} \, {\rm G}$. 
The observations are in good agreement with the predictions of the partially screened gap model, which assumes the existence of small-scale surface magnetic field structures in the polar cap region.
\end{abstract}


  \section{Introduction}

The vast majority of detected neutron stars are observed as radio pulsars \citep{2005_Manchester}. 
Almost half a century of pulsar observations have not given the definite answer to one of the most intriguing questions, namely, how the plasma that is responsible for generating the radio emission is produced. 
Although many models were proposed, no clear consensus exists over this fundamental problem until now.
There are two qualitatively quite different models that in principle can explain how a magnetosphere of a rotating highly magnetized neutron (NS) can be filled by an electron-positron plasma. 
These two models are the vacuum gap (VG) model of \cite{1975_Ruderman} and the space charge limited flow (SCLF) model of \cite{1979_Arons}.
The aim of both models is to provide a mechanism that yields a sufficiently large number of ultrarelativistic charges (electrons and positrons), which eventually create the observed radio waves. 
In both models, the plasma responsible for the generation of radio emission is produced and accelerated in a region of the open magnetic field lines above the polar cap of a pulsar. 
Over the years, both models have been modified significantly, however, no consensus has been reached on the mechanism of plasma generation.

The SCLF model assumes that surface charges can freely flow into the magnetosphere.
The density of the plasma extracted from the surface equals the Goldreich-Julian charge density at the surface, but is insufficient to screen the charge at higher altitudes \citep{2001_Hibschman}. 
In the SCLF model, pairs are created near the so-called "pair-creation front" at altitudes ranging from one meter up to 10 km \citep{2001_Harding, 2002_Harding}.
Although a connection between acceleration of plasma in the SCLF model and radio emission was not found, free particle outflow from the stellar surface is a common assumption in most of the current pulsar models.

In order to understand the phenomenon of the drifting subpulses, \cite{2003_Gil, 2006_Gil, 2006_Gil_b} developed a modification of the VG model, the so-called partially screened gap (PSG) model. 
The PSG model is based on a thermostat regulation of the thermionic release of iron ions into the inner acceleration region (IAR). 
According to \cite{2007_Medin} the cohesive energy that holds the ions within the crystal lattice is strongly dependent on the magnetic field strength at the stellar surface.
The PSG model requires the surface temperature in the polar cap region to be close to the so-called critical temperature that is defined by the surface magnetic field \citep[see][]{2007_Medin}.
X-ray observations suggest the existence of small hot spots with temperatures of a few million Kelvin \citep[see][and references therein]{2009_Becker, 2013_Szary}.
Assuming that the hot spots are actual polar caps heated by backstreaming of ultrarelativistic particles accelerated in the IAR, we can conclude that an actual area of a polar cap, $A_{\rm pc}$, is  considerably smaller than the conventional polar cap area $A_{\rm dp}\approx 6.2 \times 10^4 P^{-1} \, {\rm m^2}$ (calculated assuming a dipolar configuration of the magnetic field), here $P$ is the pulsar period in seconds.
Thus, the surface magnetic field has to be considerably stronger than the dipolar component \citep[see][for a detailed description]{2015_Szary}.

The possibility to create such field structures at the polar cap surfaces and to maintain them over the typical lifetime of pulsars has been demonstrated recently by \cite{2013_Geppert,2014_Geppert}. 
Via the Hall drift, the magnetic energy, which is stored in large-scale toroidal field configurations in deep crustal/outer core layers, is transformed into small-scale poloidal field components at the polar cap surface.

Since the predicted surface temperature of the polar cap is expected to exceed $10^6$K, X-ray observations are essential to test the model prediction, eventually leading to understanding the radio emission.
Such high temperatures at the polar cap are close to the critical temperature above which iron ions can enter the IAR. 
Therefore, the observation of $A_{\rm pc}$ -- sufficiently small and sufficiently hot -- would provide a strong support for the PSG model. 
The same is true for the ratio $A_{\rm dp}/A_{\rm pc}$. 
Given a typical dipolar surface field strength of radio pulsars $B_{\rm d}\sim 10^{11...13}$ G, magnetic flux conservation arguments require that $A_{\rm dp}/A_{\rm pc}\sim 10...1000$.

The first observations that indicated both the difference in $A_{\rm pc}$ and $A_{\rm dp}$ as well as the surface temperatures of these areas have been performed over the past 10 years, e.g. for pulsars B0943+10 \citep{2005_Zhang}, B1133+16 \citep{2006_Kargaltsev}, and J0108-14 \citep{2009_Pavlov}.
Although the X-ray photon statistics was not excellent, these results triggered more and longer observations with {\it XMM-Newton} of pulsars B0834+06 and B0826-34 that strengthened the idea of the PSG model \citep{2008_Gil}.
The discovery of synchronous mode switching in the radio and X-ray emission of PSR B0943+10 \citep{2013_Hermsen} further contributed to the increase in number of high statistics X-ray observations of pulsars.

\begin{table}[t]
    \label{tab:b1133+16}
        \caption{Observed and derived parameters of PSR B1133+16}
    \begin{center}
        \begin{tabular}{lc}
        \hline
        \hline
        & \\
        Parameter & Value \\
        & \\
        \hline
        & \\
R.A.~(J2000) \dotfill & $11^{\rm h}36^{\rm m}03\fs1829(10)$ \\
Decl.~(J2000) \dotfill & $+15^\circ 51' 09\farcs726(15)$ \\
Epoch of position (MJD) \dotfill & 51544\\
Rotation frequency, $\nu\, ({\rm Hz})$ \dotfill &  $0.84181003670065(48)$ \\
Frequency derivative, $\dot{\nu}\, {\rm  (Hz \, s^{-1})}$ \dotfill &  $-2.645070(29) \times 10^{-15}$ \\
Frequency second derivative, $\ddot{\nu}\, {\rm  (Hz \, s^{-2})}$ \dotfill &  $1.2 \times 10^{-15}$ \\
Epoch of frequency (MJD) \dotfill & 56935.339344\\
Dipolar surface magnetic field, $B_{\rm d}\, {\rm (G)}$ \dotfill & $2.13\times 10^{12}$\\
Spin-down power, $\dot{E}\,({\rm erg \, s}^{-1})$ \dotfill & $ 8.8\times 10^{31}$ \\
Characteristic age, $\tau_{\rm c}\,({\rm yr})$\dotfill & $5.04 \times 10^6$ \\
Inclination angle, $\alpha$ $(^{\circ})$ \dotfill & $52.5$\\
Opening angle, $\rho$ $(^{\circ})$\dotfill & $7.4$\\
Impact parameter, $\beta$ $(^{\circ})$\dotfill & $4.5$\\
Dispersion Measure, DM (cm$^{-3}$~pc) \dotfill & 4.86 \\
Distance from parallax, $D$ (pc) \dotfill &  357(19) \\
        & \\
        \hline
        \end{tabular}
    \end{center}
    {\bf Notes.}  
    The position and distance are from \cite{2002_Brisken}, while the frequency and its derivatives were obtained by the Jodrell Bank Pulsar Group (valid since 56190 MJD to 57681 MJD).
    The zero phase was defined between the two peaks in the radio profile.

\end{table}

The main aim of this paper is to foster the backing for the PSG model by the evaluation of another {\it XMM-Newton} observation of PSR B1133+16 (see Table \ref{tab:b1133+16} for pulsar parameters), carried out in 2014 May/June. 
With this observation, the photon statistics is significantly better than that of \cite{2006_Kargaltsev}.
Therefore, much more reliable values of the hot spot area and its respective temperature could be derived.
The fact that they fulfill the requirements of the PSG model is demonstrated in the following sections.

In order to see the correlation between the thermal states of the polar cap region and the radio emission, simultaneous observations of PSR B1133+16 in the X-ray and radio wavelength range have been performed. 
The results of these combined efforts will be presented and discussed in a forthcoming paper.

\section{Observations and data analysis}

The data were collected during five {\it XMM-Newton} observing sessions performed from 25th of May to 28th of June 2014. 
In the analysis, we have used the data obtained with the pn and MOS1/2 CCD cameras, which are part of the EPIC instrument \citep{2001_Struder, 2001_Turner}. 
In all of the observing sessions the thin optical blocking filters were used. 
The pn camera operated in the full frame mode (with timing resolution of $73\,{\rm ms}$), while both MOS operated in the small window mode (timing resolution of $0.3\,{\rm s}$).

To reduce and analyze the data collected by {\it XMM-Newton} we used the Science Analysis System (SAS ver. 14.0.0) with the latest calibration files (CCF ref. 330) and for the spectral fits we used XSPEC (ver. 12.9.0g).

\subsection{Data selection}

   \begin{figure}[t]
        \begin{center}
            \includegraphics[width=7cm]{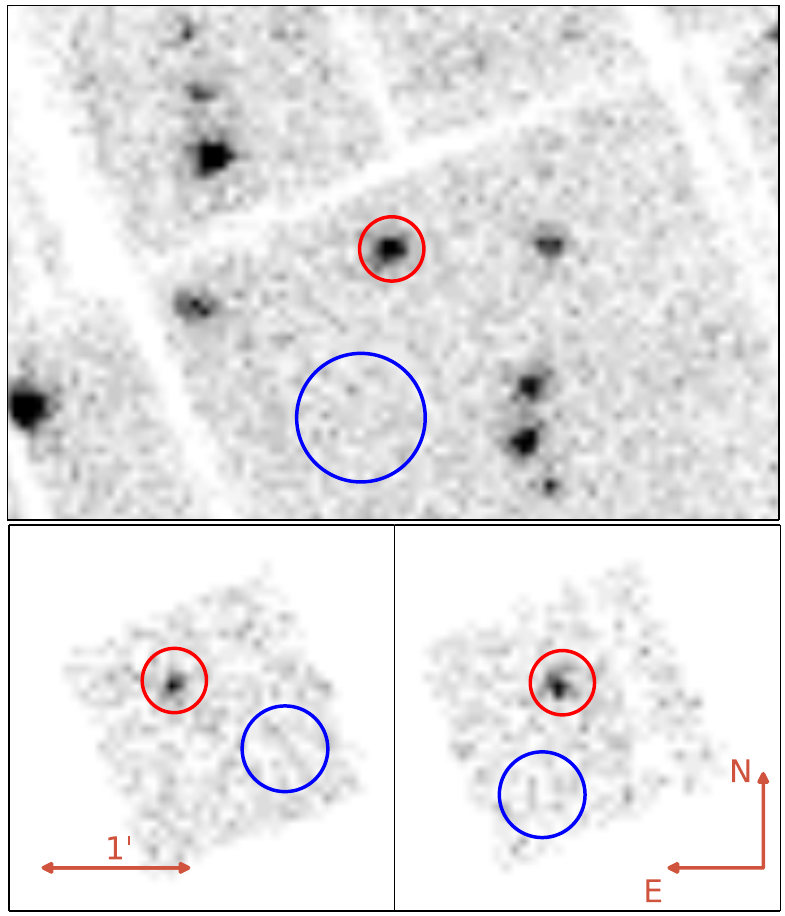}
        \end{center}
        \caption{X-ray images of the field around PSR B1133+16 in the $0.2-12\,{\rm keV}$ band. The top panel corresponds to the pn image ($6'\times4'$), while the bottom-left and bottom-right panels correspond to the MOS1  and MOS2 images ($3'\times3'$ each), respectively. The source extraction regions are marked with red and the background regions with blue circles.}
        \label{fig:image}
    \end{figure}

\begin{table*}
\centering \caption{Observation times of {\it XMM-Newton}}
\label{tab:obs}
\vspace{0.5cm}
{\small
\begin{tabular}{@{}cccrrr}
\hline
\hline
& & & & & \\
 Obs.ID & \multicolumn{2}{c}{Start/End time (UT)}   & \multicolumn{3}{c}{Observation Times (ks)} \\
        & \multicolumn{2}{c}{(YYYY-MM-DD hh-mm-ss)} & pn ()$^*$  &     MOS1  & MOS2  \\
& & & & & \\
\hline
& & & & & \\
0741140201  & 2014-05-25 12:18:42 & 2014-05-25 19:20:22 & 22.34 ({\bf 11.05}) & 24.00 ({\bf 17.12}) & 23.96 ({\bf 16.89})\\
0741140301  & 2014-05-31 11:34:51 & 2014-05-31 17:58:11 & 20.04 ({\bf 7.31}) & 21.70 ({\bf 18.40}) & 21.66 ({\bf 18.23})\\
0741140401  & 2014-06-14 07:47:26 & 2014-06-14 18:20:46 & 35.04 ({\bf 21.63}) & 36.70 ({\bf 31.49}) & 36.66 ({\bf 32.04})\\
0741140501  & 2014-06-22 07:22:13 & 2014-06-22 17:02:13 & 31.84 ({\bf 27.41}) & 33.50 ({\bf 30.36}) & 33.46 ({\bf 31.73})\\
0741140601  & 2014-06-28 10:58:52 & 2014-06-28 17:55:32 & 22.04 ({\bf 18.81}) & 23.70 ({\bf 22.43}) & 23.66 ({\bf 22.00})\\
& & & & & \\
\hline
\end{tabular}\\
$^*$ Values in parentheses correspond to the effective exposure times used in our analysis.
}
\end{table*}

\begin{figure}
        \begin{center}
            \includegraphics[width=8cm]{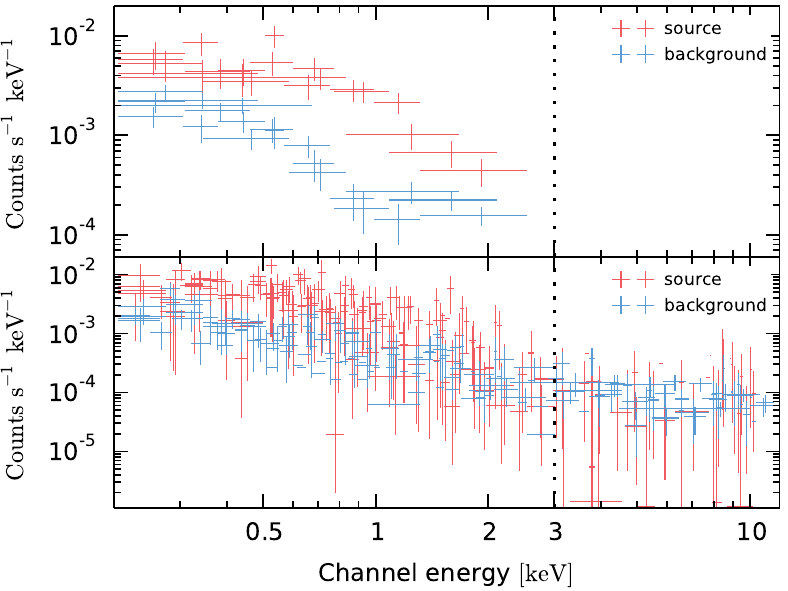}
        \end{center}
        \caption{ 
        EPIC-pn spectra from source and background regions in the $0.2-12\,{\rm keV}$ energy range.
        The top panel corresponds to the binned data, while the bottom panel corresponds to the unbinned data.
        }
        \label{fig:counts}
    \end{figure}

The total observing time was $130 \, {\rm ks}$ for pn and  $140 \, {\rm ks}$ for each of the MOS detectors.
In the analysis, we consider only photons with single and double pixel patterns for the pn detector ($\leq 4$) and up to quadrupole patterns for MOS1/2 ($\leq 12$). 
We use the {\it espfilt} task to remove soft proton contamination and any X-ray background flares. 
After removing all contaminated time intervals, the net exposure times are $86\,{\rm ks}$, $120\,{\rm ks}$ and $121\,{\rm ks}$ for the pn and MOS1/2 detectors, respectively (see Table \ref{tab:obs}). 
The source counts for the spectral and timing analysis are extracted from a circular region with a radius of $15''$ centered on the pulsar position.
The background counts are extracted from nearby source-free regions in the same CCD as the target (see Figure \ref{fig:image}).
In Figure \ref{fig:counts}, we show source and background spectra for the pn detector.
Due to the small number of high-energetic photons, we restrict the analysis to the 0.2--3 keV energy range.

\subsection{Spectral analysis}

In previous studies, spectral properties of PSR B1133+16 were derived based on 33 counts \citep{2006_Kargaltsev}.
The new {\it XMM-Newton} observation results in $378\pm 52$, $97\pm 29$, and $102\pm 33$ counts registered by the pn, MOS1 and MOS2 detectors, respectively.
For every observing session, we create source and background spectra separately for all three detectors. 
The spectra are initially binned using a constant bin size of $5\,{\rm eV}$ and $15\,{\rm eV}$ for pn and MOS1/2 detectors, respectively, to be compatible with the response files, which are based on the instrument channel width.
The response files are generated using the {\it rmfgen} and {\it arfgen} tasks.
Due to the moderate number of counts, we use both the chi-square and the Cash statistics \citep{1979_Cash} in the fitting procedures.
For the Chi-square statistics the spectral files are grouped using the {\it specgroup} task with a minimum of 20 counts in each energy bin, while for the Cash statistics with a minimum of one count in each energy bin.
Note that, to avoid adding any unnecessary error, we do not combine spectra from different observing sessions; instead, we use a joint fit to the data sets.

In order to better constrain model parameters for some fits, we fixed the hydrogen column density, $N_{\rm H}$. 
The pulsar's dispersion measure, \mbox{${\rm DM}=4.86 \, {\rm cm^{-3} \, pc}$}, corresponds to an electron column density \mbox{$N_{\rm e}=1.49 \times 10^{19}\,{\rm cm^{-2}}$}.
Assuming the typical $10\%$ ionization of the interstellar medium, we obtain \mbox{$N_{\rm H}=1.49 \times 10^{20}\,{\rm cm^{-2}}$}.

\begin{figure}
        \begin{center}
            \includegraphics[width=8.5cm]{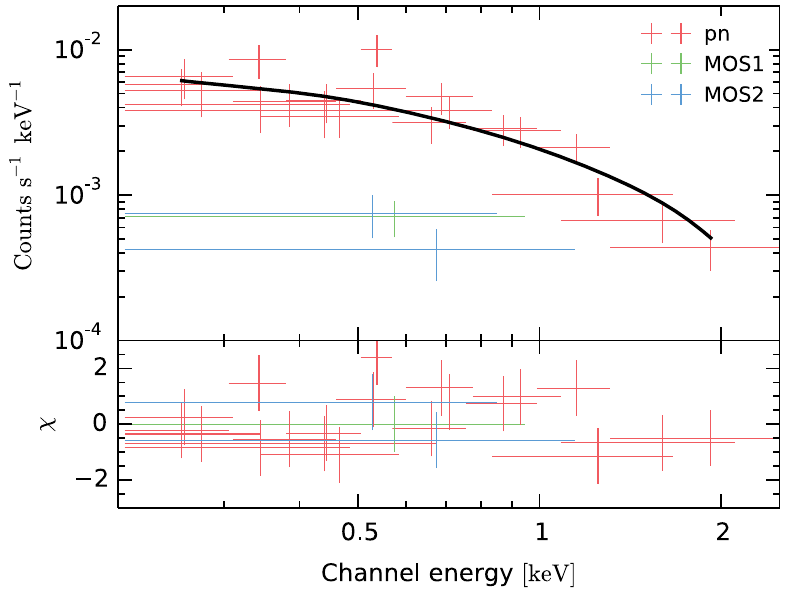}
        \end{center}
        \caption{Absorbed PL fit with fixed $N_{\rm H}=1.5\times 10^{20}\,{\rm cm^{-2}}$ for the binned pn (red), MOS1 (green), MOS2 (blue) data. In the bottom panel we show the residuals in units of sigma deviations. The black solid line corresponds to a single PL component fitted to the pn data. The models for MOS1 and MOS2 are omitted for clarity.}
        \label{fig:pl_fixed}
    \end{figure}

\begin{figure}
        \begin{center}
        \includegraphics[width=8.5cm]{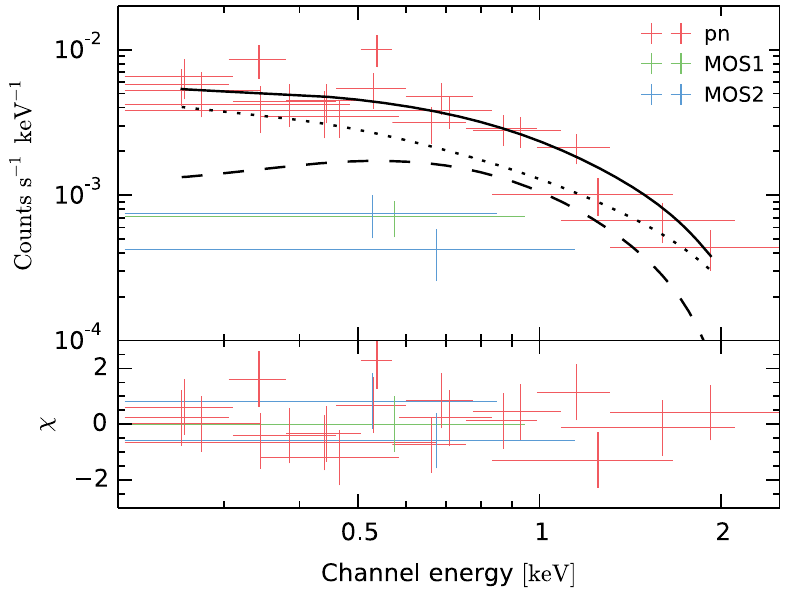}
        \end{center}
        \caption{Absorbed BB+PL fit with fixed $N_{\rm H}=1.5\times 10^{20}\,{\rm cm^{-2}}$ for the binned pn (red), MOS1 (green), and MOS2 (blue) data. In the bottom panel, we show the residuals in units of sigma deviations. The black dashed and dotted lines correspond to individual BB and PL components fitted to the pn data. The black solid line represents the total spectrum. The models for MOS1 and MOS2 are omitted for clarity.}
        \label{fig:bb_pl_fixed}
    \end{figure}

\begin{figure}
        \begin{center}
        \includegraphics[width=8.5cm]{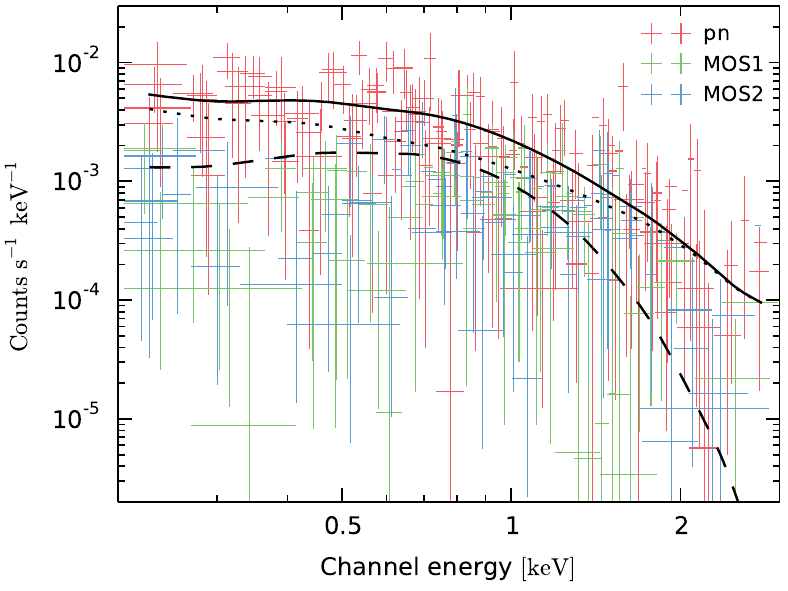}
        \end{center}
        \caption{Absorbed BB+PL fit with fixed $N_{\rm H}=1.5\times 10^{20}\,{\rm cm^{-2}}$ for the unbinned pn (red), MOS1 (green), and MOS2(blue) data. The black dashed and dotted lines correspond to individual BB and PL components fitted to the pn data. The black solid line represents the total spectrum. The models for MOS1 and MOS2 are omitted for clarity.}
        \label{fig:bb_pl_fixed2}
    \end{figure}

In Tables \ref{tab:fits} and \ref{tab:fits2}, we show the results from fitting binned and unbinned spectra, respectively.
The fit of the absorbed blackbody model (BB) with both fixed and free $N_{\rm H}$ results in poor goodness of fit with $\chi^2_{\nu}>1.3$ for binned data, and Pearson-$\chi^2_{\nu}>3$ for the unbinned data.
On the other hand, the absorbed power-law model (PL) with both fixed and free $N_{\rm H}$ describes the spectra reasonably well with $\chi^2_{\nu}=1.07$ for binned data and Pearson-$\chi^2_{\nu}=1.3$ for unbinned data (see Figure \ref{fig:pl_fixed}).
The BB+PL model fits the data equally well with $\chi^2_{\nu}<1.04$ for binned data and Pearson-$\chi^2_{\nu}=1.3$ for unbinned data (see Figures \ref{fig:bb_pl_fixed} and \ref{fig:bb_pl_fixed2}).
The spectral fits result in consistent parameters for both binned and unbinned data.
For the binned data, the observed temperature is $kT^{\infty}=0.19^{+0.06}_{-0.05} \,{\rm keV}$ with the corresponding radius of an emitting equivalent sphere $R^{\infty}_{\perp} = 17^{+18}_{-8}\,{\rm m}$, while for the unbinned data  $kT^{\infty}=0.19^{+0.04}_{-0.03}  \,{\rm keV}$, and $R^{\infty}_{\perp} = 17^{+7}_{-5}\,{\rm m}$ (see Tables \ref{tab:fits} and \ref{tab:fits2} for all spectral fit parameters).
If not stated otherwise, hereafter, we use the fit parameters for the unbinned data because they are estimated with lower statistical errors.

\begin{table*}
\caption{Parameters from spectral fits (Binned Data).}
\begin{center}
\begin{tabular}{@{}lcccccc}
\hline
\hline
  & & & & & & \\
  Parameter  & PL & PL$^{*}$ & BB & BB$^{*}$ & BB+PL & BB+PL$^{*}$ \\
  & & & & & & \\
\hline
  & & & & & & \\
$N_{\rm H}$  (10$^{20}$ cm$^{-2}$)       &   4$_{-3}^{+4}$     & $1.5$ (fixed) & $<5$    & $1.5$ (fixed)    &  $<3$  & $1.5$ (fixed) \\
Photon index                 &     $2.6^{+0.4}_{-0.3}$            & $2.3_{-0.1}^{+0.1}$  & ... &...      & $2.0_{-1.2}^{+1.3} $ & $2.4_{-0.7}^{+1.5} $ \\
$F_{PL}^{0.2-3\,{\rm keV}}$ (10$^{-15}$ erg cm$^{-2}$ s$^{-1}$) & ${8}_{-1}^{+1}$ &  ${9}_{-1}^{+1}$      & ... & ...  & ${5_{-3}^{+5}}$         &  ${7_{-3}^{+2}}$ \\
$kT^{\infty}_{}$ (keV)            &       ...             &  ...              & $0.19^{+0.01}_{-0.01}$  & $0.18^{+0.01}_{-0.01}$  & $0.18^{+0.06}_{-0.04}$  &  $0.19^{+0.06}_{-0.05}$ \\
$T^{\infty}_{}$ (MK)            &       ...             &  ...              & $2.2^{+0.1}_{-0.1}$ & $2.1^{+0.1}_{-0.1}$  & $2.1^{+0.7}_{-0.4}$   &  $2.2^{+0.7}_{-0.6}$\\
$F^{\infty}_{\rm bb}$  (10$^{-15}$ erg cm$^{-2}$ s$^{-1}$)    & ... & ...     & $7_{-3}^{+5}$ & $8_{-3}^{+5}$  & $4^{+3}_{-3}$  & $3^{+3}_{-3}$\\
$R^{\infty}_{\perp}$  (m)             &       ...          &  ...            & $25_{-3}^{+3}$ & $29_{-4}^{+4}$ & $19_{-9}^{+15}$  & $17_{-8}^{+18}$ \\
$\chi_{\nu}^2$/dof           &       1.07 / 18             & 1.07 / 19          & 1.31 / 18 & 1.38 / 19      &   1.04 / 16          & 1.00 / 17 \\
Null hypothesis probability         &    0.38     &     0.38        & 0.17 & 0.12    &  0.41         & 0.46 \\
  & & & & & & \\
 \hline
\end{tabular}
\end{center}
{\bf Notes.} Errors correspond to 1$\sigma$. {$F_{PL}^{0.2-3\,{\rm keV}}$ is the unabsorbed nonthermal flux in the $0.2-3\, {\rm keV}$ energy range, $F^{\infty}_{\rm bb}$ is the thermal bolometric flux, $T^{\infty}$ is the observed temperature, and  $R^{\infty}_{\perp}$ is the observed radius of an emitting equivalent sphere, $^*$ indicates fits with fixed $N_{\rm H}$.}
\label{tab:fits}
\end{table*}

\begin{table*}
\caption{Parameters from spectral fits (unbinned data).}
\begin{center}
\begin{tabular}{@{}lcccccc}
\hline
\hline
  & & & & & & \\
  Parameter  & PL & PL$^{*}$ & BB & BB$^{*}$ & BB+PL & BB+PL$^{*}$ \\
  & & & & & & \\
\hline
  & & & & & & \\
$N_{\rm H}$  (10$^{20}$ cm$^{-2}$)       &   $5_{-2}^{+3}$     &   $1.5$ (fixed)  &  $<5$ & $1.5$ (fixed)    &   $<2.2$  & $1.5$ (fixed) \\
Photon index                 &     $2.7^{+0.3}_{-0.2}$            & $2.3_{-0.1}^{+0.1}$  & ... &  ...      & $2.0_{-0.4}^{+0.4} $ & $2.0_{-0.4}^{+0.4} $ \\
$F_{PL}^{0.2-{3}\,{\rm keV}}$ (10$^{-15}$ erg cm$^{-2}$ s$^{-1}$) & ${8.4_{-0.8}^{+0.5}}$ &  ${9.3_{-0.5}^{+0.4}}$   & ...   & ... &    ${5_{-1}^{+3}}$         &  ${7_{-2}^{+2}}$ \\
$kT^{\infty}_{}$ (keV)            &       ...             &  ...              & $0.21^{+0.01}_{-0.01}$  & $0.21^{+0.01}_{-0.01}$ & $0.19^{+0.03}_{-0.02}$   &  $0.19^{+0.04}_{-0.03}$ \\
$T^{\infty}_{}$ (MK)            &       ...             &  ...              &  $2.5^{+0.1}_{-0.1}$   &  $2.4^{+0.1}_{-0.1}$  & $2.2^{+0.3}_{-0.3}$   &  $2.2^{+0.3}_{-0.3}$\\
$F^{\infty}_{\rm bb}$  (10$^{-15}$ erg cm$^{-2}$ s$^{-1}$)    & ... & ...     & $7^{+3}_{-2}$ & $8_{-3}^{+4}$ & $4_{-3}^{+3}$  & $3^{+3}_{-2}$ \\
$R^{\infty}_{\perp}$  (m)             &       ...          &  ...            & $15_{-2}^{+2}$ & $17_{-2}^{+2}$ & $18_{-6}^{+7}$  & $17_{-6}^{+9}$ \\
C-stat/dof           &       267 / 336             & 271 / 337     & 289 / 336      & 295 / 337      &   262 /  333        &  262 / 334 \\
P-$\chi_{\nu}^2$           &       1.30              & 1.31           & 3.24  & 3.85       &   1.31           & 1.31  \\
  & & & & & & \\
 \hline
\end{tabular}
\end{center}
{\bf Notes.} Errors correspond to 1$\sigma$. See notes in Table \ref{tab:fits}.
\label{tab:fits2}
\end{table*}

    \begin{figure*}[ht!]
        \begin{center}
        \includegraphics[]{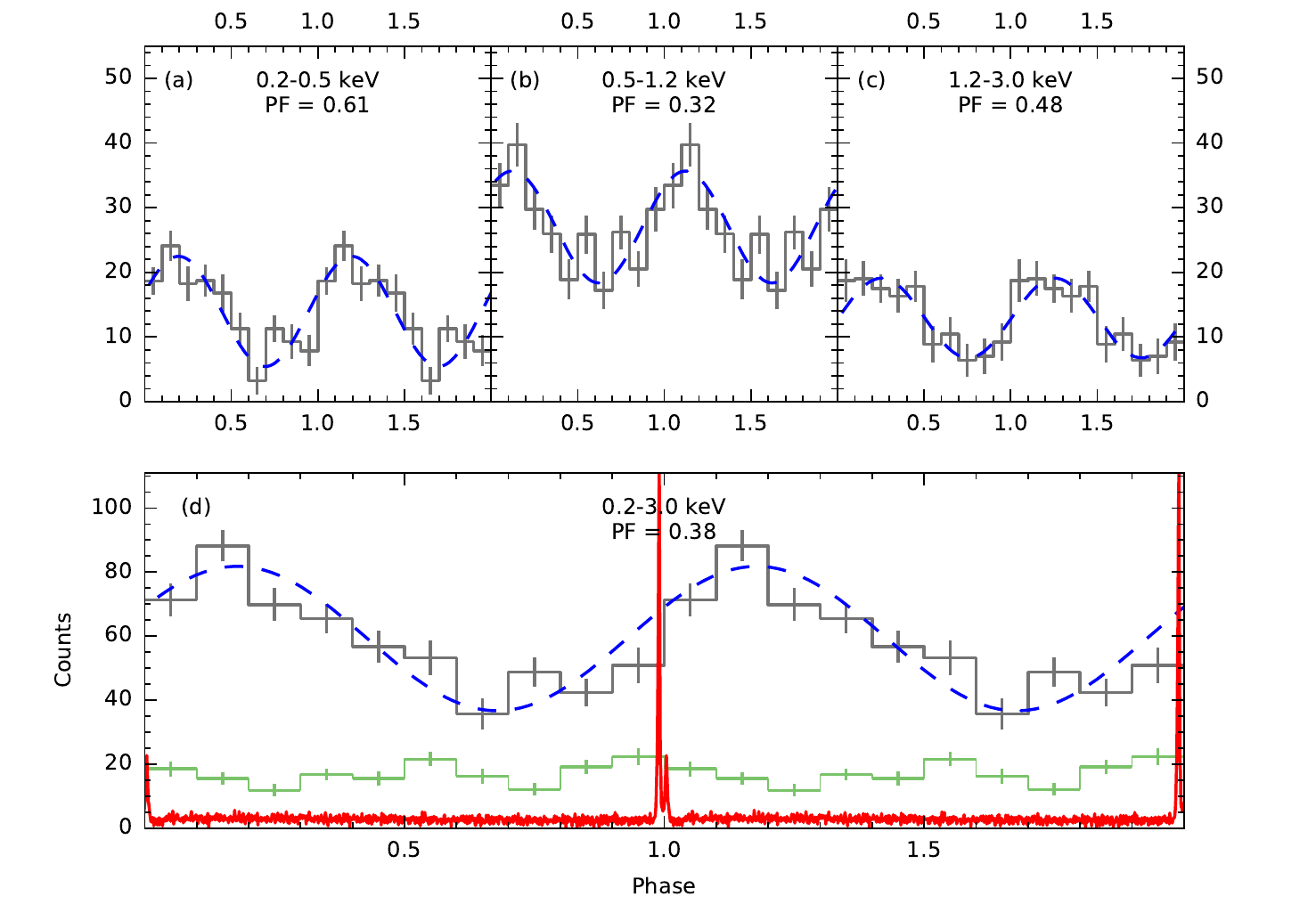}
        \end{center}
        \caption{Folded pn + MOS1/2 source light curves in four different energy ranges. The blue dashed lines show fits with sinusoidal functions, while the red solid line shows the Effelsberg radio profile at $4.858 \, {\rm GHz}$ in arbitrary units. The green line corresponds to the background level. Note that the light curves are corrected for background. Two phase cycles are shown for clarity.}
        \label{fig:lightcurve}
    \end{figure*}

\clearpage

\subsection{Timing analysis}

The source light curves are extracted from a circular region with a radius of $15''$ centered on the pulsar position, while the background light curves are extracted from nearby source-free regions.
We use the {\it barycen} task and ephemeris of PSR B1133+16 given in Table \ref{tab:b1133+16} to convert photon arrival times to the Barycentric Dynamical Time.
Both absolute corrections (vignetting, bad pixels, chip gaps, quantum efficiency, etc.) and relative corrections (background counts, dead times, etc.) are performed with the {\it epiclccorr} task.
The pulse phases of the counts are computed using the {\it phasecalc} task with ephemeris of the pulsar (see Table \ref{tab:b1133+16}).
The uncertainties in pulsar ephemeris result in a negligible delay of  $0.01^{\circ}$.
In Figure \ref{fig:lightcurve}, we show the folded pulse profiles in four energy ranges: $0.2-0.5\,{\rm keV} $, $0.5-1.2\,{\rm keV} $, $1.2-3\,{\rm keV} $, and $0.2-3\,{\rm keV}$.
The light curves are broadly sinusoidal and are characterized with a delay of about $70^{\circ}$ from the radio main pulse (see the red solid line in panel (d) in Figure \ref{fig:lightcurve}).
Assuming that X-ray pulsations are connected with the polar cap radiation, we can estimate the polar cap dislocation from the magnetic axis: $d=\delta R \sin \alpha $, where $\delta$ is the delay in radians, $R$ is the neutron star radius, and $\alpha$ is the inclination angle.
Using the inclination angle of the pulsar, $\alpha=52.5^{\circ}$, and $R=10\,{\rm km}$ we get the polar cap dislocation $d \approx 9.7 \,{\rm km}$.
Such a large dislocation is not possible assuming crust anchored small-scale magnetic anomalies and would require a non-dipolar global magnetic field.
Thus, the phase-shift between X-ray maximum and radio peak suggests a nonthermal origin of the X-ray pulsed emission.
In Table \ref{tab:lightcurve} we show the properties inferred from the light curves in different energy ranges.
The uncertainties in X-ray/radio delays and pulsed fractions were estimated based on 100 light curves constructed using different good time intervals. 
The difference in the good time intervals is due to the randomization of photons' energy and arrival times according to the energy/time resolution.
Interestingly, in the energy range of $0.5-1.2\,{\rm keV}$, where the influence of the BB component on the spectrum is strongest (see Figures \ref{fig:bb_pl_fixed} and \ref{fig:bb_pl_fixed2}), both the delay from the radio peak and the pulsed fraction are lower.
This suggests the presence of the BB component with a considerably smaller pulsed fraction and smaller delay from the radio peak than the PL component.

    \begin{table}[ht!]
        \caption{Properties of X-Ray Light Curves of PSR B1133+16.}
        \begin{center}
        \begin{tabular}{lccc}
        \hline
        \hline
        & & \\
        Energy Range & Counts & X-Ray/Radio Delay  & Pulsed Fraction  \\
        $\rm (keV)$ & & $\rm (^{\circ})$ & $(\%)$ \\
        & & & \\
        \hline
        & & & \\
        $0.2-0.5$ & $398 \pm 71$ & $70 \pm 8$ & $61 \pm 5$ \\
        & & & \\
        $0.5-1.2$ & $442 \pm 73$ & $44 \pm 9$ & $32 \pm 4$ \\
        & & & \\
        $1.2-3$ & $299 \pm 64$ & $92 \pm 7$ & $48 \pm 3$ \\
        & & & \\
        $0.2-3$ & $581 \pm 95$ & $64 \pm 7$ & $38 \pm 3$ \\
        & & & \\
        \hline
        \end{tabular}
        \end{center}
        {\bf Notes.}  The individual columns are as follows: (1) considered energy range, (2) source counts in the considered energy range, (3) delay between X-ray and radio profiles, and (4) pulsed fraction defined as the amplitude of the sinusoid divided by the average value. The uncertainties correspond to one sigma.
        \label{tab:lightcurve}
    \end{table}

\section{Physical implications for IAR}

Under the assumption that the thermal component of the pulsar X-ray emission comes from the actual hot polar cap, the {\it XMM-Newton} data allows us to constrain the pulsar polar cap physics and to test the predictions of the PSG model.

\subsection{Partially Screened Gap} 
\label{sec:psg}
The PSG model was introduced by \cite{2003_Gil, 2006_Gil, 2006_Gil_b} in order to explain the observational rates of drifting subpulses. 
In PSG, the full vacuum potential drop is largely reduced by thermal iron ions $\rm Fe^{56}$ thermally ejected from the hot surface of the polar cap, therefore reducing drift rates that result from the VG model. 
The potential drop of PSG can be described as
        \begin{equation}
            \Delta V=\eta\Delta V_{{\rm max}},\label{eq:deltaV}
        \end{equation}
where $\Delta V_{{\rm max}}$ is the potential drop in a VG and $\eta$ is the screening factor. The screening factor is
characterized by the ratio of the density of iron ions $\rho_{\rm i}$ to the Goldreich-Julian co-rotational density $\rho_{\rm GJ}$ as
follows.
        \begin{equation}
            \eta=1-\rho_{{\rm i}}/\rho_{{\rm GJ}}.
        \end{equation}
\cite{2007_Medin} showed that, for the observed surface temperatures (a few million Kelvin), electrons can easily escape from the stellar
surface, thereby completely screening the acceleration potential above the polar cap. 
Thus, for neutron stars with ${\bf \Omega}\cdot{\bf B}>0$, we do not expect the formation of any acceleration region above the polar cap, here $\bf \Omega$ is the neutron star rotation axis.
In the case of neutron stars with ${\bf \Omega}\cdot \mathbf{B}<0$, positive charges (iron ions) are responsible for the screening of the potential drop. 
Since the density of the iron ions in the neutron star crust is many orders of magnitude larger than the co-rotational charge density, only a small fraction is enough to completely screen the accelerating potential. 
In order to form PSG, the surface temperature $T_{\rm s}$ needs to be below the so-called critical value $T_{\rm crit}$, i.e. the temperature at which the density of the outflowing thermal ions equals the Goldreich-Julian co-rotational density. 
The cohesive energy of iron ions is mainly defined by the strength of the magnetic field. 
By fitting to the numerical calculations of \cite{2007_Medin}, we can find the dependence of the critical temperature, $T_{\rm crit}$, on the pulsar parameters
        \begin{equation}
            T_{{\rm crit}}\approx 2 \times10^{6} \, B_{14}^{0.75} \, {\rm K}, \label{eq:t_crit}
            \end{equation}
where $B_{14}=B_{{\rm s}}/\left(10^{14}\,{\rm G}\right)$, $B_{\rm s}=b B_{\rm d}$ is a surface magnetic field, and $b=A_{\rm dp}/A_{\rm
pc}$ (applicable only if the polar cap radiation is revealed with $A_{\rm pc} < A_{\rm dp}$).

\begin{table*}[ht!]
\caption{Polar Cap Properties of PSR B1133+16.}
\begin{center}
\begin{tabular}{lcccccc}
\hline
\hline
 &  &  &  &  & & \\
Radiation Source  &  $\left < \cos i \right >$  & $\left < f \right >$  & $R_{\rm pc}$ & $T_{\rm s}$ & $\log L_{\rm bb} $ & Pulsed Fraction\\
 &  &  & $\rm (m)$ & $\rm (MK)$  & $\rm ( erg\,s^{-1})$ & ($\%$)\\
 &  &  &  &  & & \\
\hline
 &  &  &  &  & & \\
Primary spot & 0.40 & 0.61 & $16^{+9}_{-6}$ & $2.9^{+0.6}_{-0.4}$ & $28.5^{+0.4}_{-0.4}$ & 64 \\
 &  &  &  & & & \\
Antipodal spot & 0.07 & 0.25 & $26^{+14}_{-10}$ & $2.9^{+0.6}_{-0.4}$ & $28.9^{+0.4}_{-0.4}$ & 100 \\
 &  &  &  &  & & \\ 
Two spots & 0.48 & 0.86 & $14^{+7}_{-5}$ & $2.9^{+0.6}_{-0.4}$ & $28.4^{+0.4}_{-0.4}$ & 9 \\
 &  &  &  &  & & \\
\hline
\end{tabular}
\end{center}
{\bf Notes.}  The individual columns are as follows: (1) assumed source of radiation, (2) time-averaged cosine of the angle between the magnetic axis and the line of sight, (3) flux correction factor (including gravitational bending of light), (4) radius of the polar cap, (5) temperature of the polar cap, and (6) Bolometric luminosity. The gravitational bending effect was calculated using $M=1.4\,{\rm M}_{\odot}$ and $R=10\,{\rm km}$.
 \label{tab:fractions}
\end{table*}

\subsection{Observations of the polar cap}

The blackbody fit to the observed spectrum of a pulsar allows us to obtain the redshifted (measured by a distant observer) effective temperature $T^{\infty}$ and redshifted total bolometric flux $F^{\infty}$.
The unredshifted (actual) parameters can be estimated by taking into account the gravitational redshift, $g_{{\rm r}}=\sqrt{1-2GM/Rc^{2}}$, determined by the neutron star mass $M$ and radius $R$, here $G$ is the gravitational constant.
The actual effective temperature and actual total bolometric flux can be estimated as $T_{\rm s}=g_{{\rm r}}^{-1}T^{\infty}$, $F_{\rm bb}=g_{{\rm r}}^{-2}F^{\infty}$ \citep[see, e.g.,][]{2007_Zavlin}.
If the distance to the neutron star, $D$, is known, we can use $T$ and $F$ to calculate the size of the radiating region.
Assuming that the radiation is isotropic (e.g. radiation from the entire stellar surface), we can estimate the radius as $R_{\perp} = g_{\rm r} R_{\perp}^{\infty} = D\sqrt{g_{\rm r}^2 F^{\infty}/(\sigma T^{\infty4})}$, where $\sigma \approx 5.6704\times10^{-5}{\rm \, erg\, cm^{-2}\, s^{-1}\, K^{-4}}$ is the Stefan--Boltzmann constant.

If we assume that the radiation comes from the hot spot (or spots) on the stellar surface, the modeling of thermal radiation is more complicated.
The following factors must be taken into consideration: the time-averaged cosine of the angle between the magnetic axis and the line of sight $\left < \cos i \right >$, the gravitational bending of light, as well as the number of the radiating spots, i.e.  whether the radiation comes from two spots (e.g., from opposite poles of the star) or from one hot spot only. 
The observed radius of the radiating spot is influenced by the geometrical factor $f$, which
depends on the following angles: $\alpha$ between the spin and magnetic axes and $\zeta$ between the line of sight and the spin axis, as well as on $g_{{\rm r}}$ and whether the radiation comes from the star's two opposite poles or from a single hot spot only: $R_{{\rm
pc}}=g_{{\rm r}}f^{-1/2}R_{{\perp}}^{\infty}$.

Since the radius of a neutron star is only a few times larger than the Schwarzschild radius, a strong gravitational field just above the
stellar surface causes the bending of light. For a Schwarzschild metric, we can calculate the observed flux fraction from the primary and
antipodal spots with respect to the maximum possible flux that is observed when the primary spot is viewed face-on
\citep{2002_Beloborodov}:
    \begin{equation}
        \begin{array}{ccccc}
        f_{1} & = & \cos i\left(1-\frac{r_{{\rm g}}}{R}\right)+\frac{r_{{\rm g}}}{R}, &{\rm if}& \cos i >-\frac{r_{{\rm g}}}{\left(R-r_{{\rm g}}\right)}\\
        f_{2} & = & -\cos i \left(1-\frac{r_{{\rm g}}}{R}\right)+\frac{r_{{\rm g}}}{R} &{\rm if}& \cos i <\frac{r_{{\rm g}}}{\left(R-r_{{\rm g}}\right)},\\
        \end{array}
    \end{equation}
here $r_{{\rm g}}=2GM/c^{2}$ is the Schwarzschild radius and $\cos i$ is the cosine of the angle between the magnetic axis and the line of
sight \citep[see, e.g.,][]{2013_Szary}. Note that when $-r_{{\rm g}}/\left(R-r_{{\rm g}}\right)<\cos i<r_{{\rm g}}/\left(R-r_{{\rm
g}}\right)$ both spots are seen and the observed flux fraction is $f_{{\rm min}}=f_{1}+f_{2}=2r_{{\rm g}}/R$.

\subsection{PSR B1133+16}

    \begin{figure}
        \begin{center}
        \includegraphics{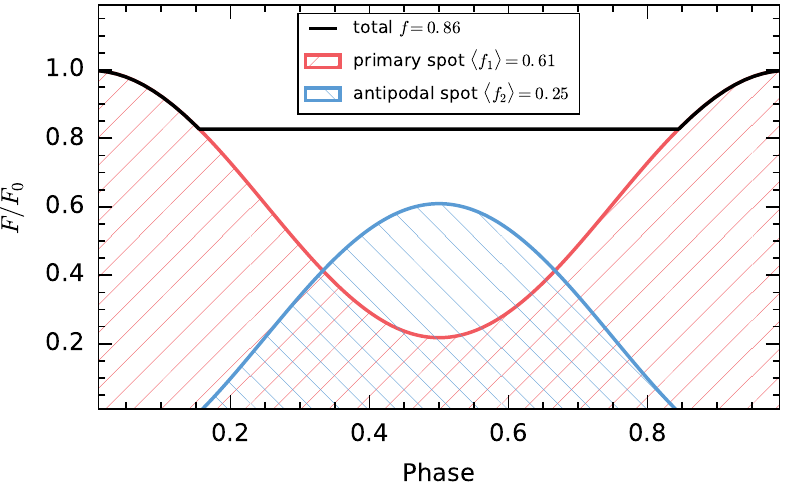}
        \end{center}
        \caption{Observed flux fraction $f$ as a function of the rotation phase for PSR B1133+16.
The following parameters were used: $\alpha=52.5^{\circ}$, $\beta=4.5^{\circ}$, $M = 1.4M_{\odot}$, $R=10\,{\rm km}$.}
        \label{fig:flux_fraction}
    \end{figure}

\begin{figure}
        \begin{center}
        \includegraphics[width=8.5cm]{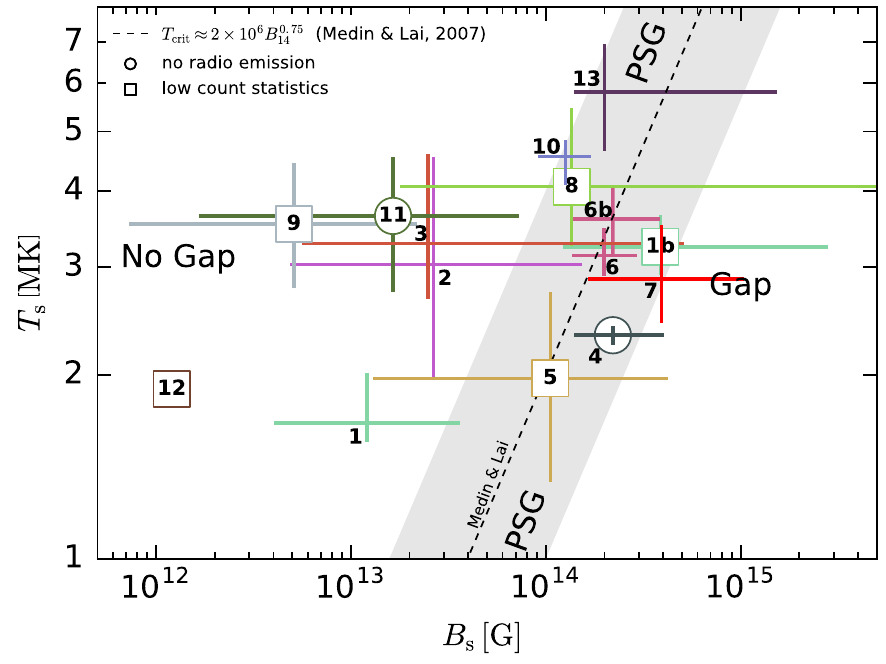}
        \end{center}
        \caption{Diagram of the surface temperature, $T_{\rm s}$, vs. the surface magnetic field, $B_{\rm s}$. The dashed line represents the dependence of $T_{{\rm crit}}$ on $B_{s}$ according to \citet{2007_Medin}, and the gray region corresponds to uncertainties in the theoretical predictions. Error bars correspond to $1 \sigma$. Pulsars: (1) J0108-1431, (2) B0355+54, (3) B0628-28, (4) J0633+1746, (5) B0834+06, (6) B0943+10, (7) B1133+16, (8) B1451-68, (9) B1719-37, (10) B1929+10, (11) J2021+4026, (12) J2043+2740, (13) B2224+65} \citep[see][and references therein]{2009_Becker, 2013_Szary}.
        \label{fig:t6_b14}
    \end{figure}

As shown in the previous section, knowing the pulsar geometry \citep{1988_Lyne, 1997_Kijak}, we can estimate the observed flux depending on the source of thermal radiation.
In Figure \ref{fig:flux_fraction} we present the observed flux fraction as a function of the rotational phase for PSR B1133+16.
If the two polar caps have the same properties (temperature and size), the gravitational bending increases contribution of radiation from the antipodal spot from about $15\%$ to $30\%$.
In Table \ref{tab:fractions}, we present flux fractions and corresponding polar cap properties calculated assuming different sources of radiation, i.e. the primary polar cap only, the antipodal polar cap only, or the two polar caps.
In the PSG model, an actual temperature at the polar cap depends on the magnetic field strength in that region, and thus on the non-dipolar configuration. 
In general, there is no reason that the configuration of the magnetic field at two opposite polar caps should be similar. 
Therefore, for the geometry of PSR B1133+16 the best solution would be to use a three component fit: two blackbody and power-law components. 
Unfortunately, the count statistics is not enough to perform such a fit.
On the other hand, we can use the pulsed fraction information from the timing analysis to answer the question of the origin of the thermal emission.
The pulsed fraction in the $0.5-1.2\,{\rm keV}$ energy range decreases by about $10 \%$ with respect to the neighboring energy ranges (see Table \ref{tab:lightcurve}).
This suggests that the pulsed fraction of thermal emission is below $30 \%$.
The predicted pulsed fractions of the thermal component (see Table \ref{tab:fractions}) suggest that thermal radiation originates from both primary and antipodal polar caps.
We found that regardless of the assumed source of radiation (primary, antipodal, or two hot spots), the properties of blackbody radiation, $T_s \approx 2.9\,{\rm MK}$, $R_{\rm pc}\approx 14\,{\rm m}$, and the derived surface magnetic strength at the pole, $B_{\rm s}\approx 3.9\times 10^{14}\,{\rm G}$, are in full agreement with the predictions of the PSG model (see Figure \ref{fig:t6_b14}).

\section{Conclusions}





The spectrum of B1133+16 can be well described by a single PL component (see Table \ref{tab:fits2}).
The $\sim 70^{\circ}$ misalignment between X-ray and radio peaks confirms a nonthermal origin of the X-ray pulsed emission.
In the energy range of $0.5-1.2\,{\rm keV}$, both the delay from the radio peak and the pulsed fraction are significantly lower than in the neighboring energy ranges.
This suggests the presence of an additional component, namely the thermal emission with considerably smaller pulsed fraction than the nonthermal component.
To produce thermal emission with a relatively low pulsed fraction, radiation has to originate from two polar caps (see Figure \ref{fig:flux_fraction}).
The blackbody spectral fit indicates a surface temperature of $T_{\rm s}=2.9$MK and a radius of the heated polar caps of $R_{\rm pc}\approx 14$m.
Relating this radius to the radius given by the last open field line of the dipolar magnetic field component $R_{\rm dp}\approx 130$m, the local polar cap surface magnetic field strength $B_{\rm s} \approx 3.9 \times 10^{14}$G.

The main conclusion from the {\it XMM-Newton} observations of PSR B1133+16 can be drawn from Figure \ref{fig:t6_b14}.
One can see from the figure that most rotation powered normal radio pulsars line within 1$\sigma$ from the Medin-Lai critical value.
The only exceptions are pulsars either with a number of photons that is too low to reliable fit their spectra (e.g., B0834+06 (No. 5), B1451-68 (No. 8), B1719-37 (No. 9), and J2043+2740 (No. 12)) or without any detected radio emission (e.g., J2021+4026 (No. 11)).
It indicates the possibility to form an accelerating potential gap above the polar cap.
It can generally be stated this recent {\it XMM-Newton} observation of PSR B1133+16 supports the ideas on which the PSG model of the inner accelerating gap physics are based.

On the other hand, however, we have to note the large systematic uncertainty in the analysis of X-ray data of pulsars.
The X-ray spectrum of pulsars can be interpreted in the framework of a completely different family of models, i.e. the neutron star atmosphere models.
In such models the thermal component is emitted from a very thin, $\sim 0.1-10\,{\rm cm}$, neutron star surface layer.
Light-element atmosphere models result in a temperature $T_{\rm atm}$ that is significantly lower than the blackbody temperature $T_{\rm bb}$, with a typical ratio of $T_{\rm bb} / T_{\rm atm} \sim 2-3$ and considerably larger emitting area, by a factor of $\sim 50-200$  \citep[see, e.g.,][]{1996_Pavlov, 2001_Pavlov, 2002_Pons, 2006_Adelsberg}.
The atmosphere models are used to study the thermal evolution of neutron stars, to constrain the equation of state and composition of the superdense matter in the neutron star interior and other issues beyond the scope of this paper.
Note that if the outer layers of a neutron star consist of light elements, the formation of PSG is not possible due to the low cohesive energy of the ions \citep{2007_Medin}.
In such a case, SCLF can be responsible for plasma generation and acceleration.

The X-ray data of PSR B1133+16, whose analysis is presented here, were obtained in an XMM-observation carried out simultaneously with radio observations of this pulsar that employed the GMRT, Effelsberg, Kunming, and the LOFAR radio telescopes. 
In a forthcoming paper we will present a detailed timing analysis of radio and X-ray data and discuss further conclusions.

\acknowledgments

This paper is dedicated to the memory of our colleague, the late Professor Janusz Gil, who pioneered the PSG model and many other seminal works in Pulsar Astrophysics.
This work is supported by National Science Centre Poland under grants DEC-2012/05/B/ST9/03924 and DEC-2013/09/B/ST9/02177 and by the Netherlands Organisation for Scientific Research (NWO) under project "CleanMachine" (614.001.301). RXX acknowledges supports from NSFC (grant No. 11673002 and U1531243). We thank the Jodrell Bank Pulsar Group for the updated ephemeris obtained with the Lovell Telescope funded by the STFC. We thank the anonymous referee for constructive comments that helped us to improve the paper significantly. Special thanks to Jason Hessels and Benjamin Stappers for providing us with the necessary data to perform adjustment of radio and X-ray observations.

\end{document}